\def\answ{b }
\input harvmac.tex\overfullrule=0pt
\input mssymb.tex
\input labeldefs.tmp
\writedefs
\def\Title#1#2{\rightline{#1}
\ifx\answ\bigans\nopagenumbers\pageno0\vskip1in%
\baselineskip 15pt plus 1pt minus 1pt
\else
\def\listrefs{\footatend\vskip
1in\immediate\closeout\rfile\writestoppt
\baselineskip=14pt\centerline{{\bf References}}\bigskip{\frenchspacing%
\parindent=20pt\escapechar=` \input refs.tmp\vfill\eject}\nonfrenchspacing}
\pageno1\vskip.8in\fi \centerline{\titlefont #2}\vskip .5in}
 
scaled\magstep3
 
scaled\magstep3
 
scaled\magstep3
 
scaled\magstep3
 
scaled\magstep3
\ifx\answ\bigans\def\tcbreak#1{}\else\def\tcbreak#1{\cr&{#1}}\fi
%
%

\def\inbar{\,\vrule height1.5ex width.4pt depth0pt}
\def\IB{\relax{\rm I\kern-.18em B}}
\def\IC{\relax\hbox{$\inbar\kern-.3em{\rm C}$}}
\def\IP{\relax{\rm I\kern-.18em P}}
\def\IR{\relax{\rm I\kern-.18em R}}

\def\a{\alpha}

\def\cc{{\bf c}}
\def\sn{\mathop{\rm sn}\nolimits}
\def\cn{\mathop{\rm cn}\nolimits}
\def\dn{\mathop{\rm dn}\nolimits}
\def\dc{\mathop{\rm dc}\nolimits}
\def\tr{\mathop{\rm tr}\nolimits}

\def\Re{\mathop{\rm Re}\nolimits}
\def\Im{\mathop{\rm Im}\nolimits}

\def\Dsl{\,\raise.15ex\hbox{/}\mkern-13mu D}
\def\Hsl{\,\raise.15ex\hbox{/}\mkern-12.5mu H}
\def\Lsl{\,\raise.15ex\hbox{/}\mkern-13mu L}
\def\hsl{\raise.15ex\hbox{/}\kern-.57em h}
\def\omegasl{\raise.15ex\hbox{/}\kern-.57em\omega}
\def\der{\partial}
\def\e#1{{\rm e}^{#1}}
\def\atr#1{\left<{\tr\over N}{#1}\right>}
%
\input epsf

\long\def\fig#1#2#3{%
\xdef#1{\the\figno}%
\writedef{#1\leftbracket \the\figno}%
\midinsert%
\parindent=0pt\leftskip=1cm\rightskip=1cm\baselineskip=11pt%
\centerline{\epsfbox{#3}}
\vskip 8pt\ninepoint%
{\bf Fig. \the\figno:} #2%
\endinsert%
\goodbreak%
\global\advance\figno by1%
}
\Title{\vbox{\baselineskip12pt
\hfill{\vbox{
\hbox{LPTENS-98/27} 
}}}}
{\vbox{\centerline{Two-Matrix Model with $ABAB$ Interaction }}}
\bigskip
\bigskip
\centerline{Vladimir A. Kazakov {\it and}\/ Paul Zinn-Justin}
\bigskip
\centerline{ Laboratoire de Physique Th\'eorique de}
\centerline{l'Ecole Normale Sup\'erieure\footnote*{
Unit\'e Propre du
Centre National de la Recherche Scientifique,
associ\'ee \`a l'Ecole Normale Sup\'erieure et \`a
l'Universit\'e de Paris-Sud.}}
\centerline{24 rue Lhomond, 75231
Paris Cedex 05, France}
\bigskip
\noindent
Using recently developed methods of character expansions we solve
exactly in the large 
$N$ limit a new two-matrix model of hermitean matrices $A$ and $B$
with the action 
$S={1\over 2}(\tr A^2+\tr B^2)-{\alpha\over 4}(\tr A^4+\tr
B^4)-{\beta\over 2} \tr(AB)^2$. 
This model can be mapped onto a special case of the 8-vertex
model on dynamical planar graphs. The solution is parametrized
in terms of elliptic functions. A phase transition is found:
the critical point is a conformal field theory with central
charge $\cc=1$ coupled to 2D quantum gravity.

\Date{July 1998}\def\rem#1{}

\newsec{Introduction}

Matrix models a proven to be a powerful tool in the study of various mathematical 
and physical problems, such as random geometry and enumeration of graphs, 
strings and two dimensional quantum gravity,
chaos and mesoscopic systems, statistical mechanics of spins on random
lattices etc. 

The list of solvable matrix models is not very long.
Solvability is usally connected to reducing the number of relevant
degrees of freedom
from $\sim N^2$ to $\sim N$, for matrices of size $N\times N$.
Two basic and very important examples are
given by the one matrix model with the partition function:
\eqn\ONEMM{{\rm Z_{1MM}}(\alpha)=
\int dM\, \exp N\left[-{1\over 2}\tr M^2+{\alpha\over 4}
\tr M^4\right]
}
and the two matrix model:
\eqn\TWOMM{{\rm Z_{2MM}}(\alpha,\beta)=
\int\!\!\!\int dA\, dB\, \exp N\left[-{1\over 2}(\tr A^2+\tr B^2)
+{\alpha\over 4}(\tr A^4+\tr B^4)+ {\beta\over 2} \tr(AB)\right]
}
where the integrals go over the $N\times N$ hermitean matrices $M$,
$A$, $B$. We consider here the frequently used quartic non-linearities
in the matrix potentials, although one could have a more general potential.
 
The large $N$ limit of these models is perturbatively
equivalent to the planar (or spherical) Feynman 
graph expansion and describes interesting physical systems. 
The first one is related to the 
problems of enumeration of planar graphs 
\ref\BIPZ{E.~Br\'ezin, C.~Itzykson, G.~Parisi and J.-B.~Zuber,
{\it Commun. Math. Phys.} 59 (1978), 35.}
and was proposed as a model of pure 2D gravity
\nref\DAVID{F.~David, {\it Nucl. Phys.} B257 (1985), 45.}%
\nref\VOL{V.A.~Kazakov, {\it Phys. Lett.} B150 (1985), 282.}%
[\xref\DAVID,\xref\VOL]; the second, solved in the planar limit in 
\ref\IZ{C.~Itzykson and J.-B.~Zuber, {\it J. Math. Phys.} 21 (1980), 411.}
and \ref\MEHTA{M.L.~Mehta, {\it Comm. Math. Phys.}
79 (1981), 327}
describes the Ising model on random dynamical planar graphs
\ref\KAZBOUL{V.A.~Kazakov, {\it Phys. Lett.} A119 (1986), 140\semi
D.V.~Boulatov and V.A.~Kazakov, {\it Phys. Lett.} B214 (1988), 581.}.
It is clear that the exact solvability of any natural generalizations
or modifications of these models could  be very useful in various physical and
mathematical applications. 

We present in this paper a very natural solvable generalization of the
two matrix model 
given by the following integral:
\eqn\defZ{{\rm Z}(\alpha,\beta)=
\int\!\!\!\int dA\, dB\, \exp N\left[-{1\over 2}(\tr A^2+\tr B^2)
+{\alpha\over 4}(\tr A^4+\tr B^4)+{\beta\over2} \tr(AB)^2\right]
}
The only difference with the conventional
two-matrix model \TWOMM\ resides in the last term 
in the matrix potential. This ``little'' modification
will turn out to be very important.

\nref\HC{Harish~Chandra, {\it Amer. J. Math.} 79 (1957), 87.}
First of all, this new model cannot be solved by the same method as
the old matrix models:
the Itzykson--Zuber--Harish~Chandra [\xref\HC,\xref\IZ]
formula is not applicable here to reduce it to an eigenvalue problem.
In order to avoid this problem,
we shall apply a character expansion method, which was worked out in 
connection with the matrix models in the papers
\ref\IDiF{P.~Di~Francesco and C.~Itzykson, {\it Ann. Inst. Henri
Poincar\'e} 59, no. 2 (1993), 117.},
\ref\KaSW{V.A.~Kazakov, M.~Staudacher and T.~Wynter,
{\it Commun. Math. Phys.} 177 (1996), 451; 179 (1996), 235;
{\it Nucl. Phys.} B471 (1996), 309.}, \ref\KoSW{
I.~Kostov, M.~Staudacher and T.~Wynter,
{\it Commun. Math. Phys.} 191 (1998), 283.}
where it was successfully applied to the investigation
of 2D $R^2$ gravity \KaSW\ and the enumeration of branched coverings
\KoSW.
The basic point of this method is the reduction of the degrees of freedom 
(from $\sim N^2$ to $\sim N$) in terms of
highest weights of representations in the character expansion. 
It is a sort of Fourier transform for the matrix variables. 
The large $N$ limit allows to apply the 
saddle point approximation for the highest weight distribution 
(one looks for the most probable Young tableau).   
Here, following the idea of \ref\PZJ{P.~Zinn-Justin,
{\it Commun. Math. Phys.} 194 (1998), 631.}, we shall use a double
saddle point on both eigenvalues and highest weights.

Second, the planar graph expansion for the model \defZ\ corresponds
to a different statistical 
mechanical system than the Ising spins on a random dynamical lattice 
of the conventional 
two matrix model \TWOMM. The $\phi^4$-type planar diagrams of the
model consist from the 
intersecting and self-intersecting closed paths of 2 different colours
(see fig. \diag). 
The colouring contributes to the entropy and thus describes
some spin degrees of freedom. 
\fig\diag{A typical graph in the expansion of the
two matrix model with $ABAB$ interaction. The two matrices
($A$, $B$) are represented by two colours (green, red).}{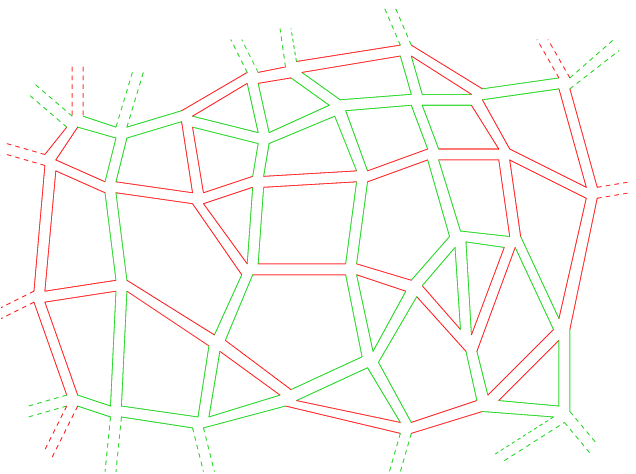}

Alternatively, this statistical model model can be mapped on a 
special case of the {\it 8-vertex model} on random dynamical graphs. 
To see it let us set $X=A+iB$; $X$ is an 
arbitrary complex matrix, and the model can be recast as a one-matrix model:
\eqn\onecmm{
{\rm Z}(b,c,d)=\int dXdX^\dagger\,
\exp N\left[-{1\over 2}\tr(XX^\dagger)
+b\tr(X^2 X^{\dagger 2})
+{c\over2}\tr(XX^\dagger)^2
+{d\over4}\tr(X^4+X^{\dagger 4})
\right]
}
Since $X$ is complex, the propagators carry arrows, and the three
vertices reproduce the well-known configurations of the 8-vertex model.
Here, the three corresponding weights $b$, $c$, $d$ are not independent,
since they are related to $\alpha$ and $\beta$ by\foot{To have
the three independent constants, one would have to
introduce an additional interaction $\gamma \tr(A^2B^2)$.}:
$$\eqalign{
b&={\alpha+\beta\over 8}  \cr
c=d&={\alpha-\beta\over 8}\cr
}$$

On a regular lattice, the 8-vertex model is critical when the weight
$d$ goes to zero, so that we are left with the {\it 6-vertex model}.
In the standard parametrization of the 6-vertex model (see for example
\ref\ZUB{J.B.~Zuber in Les Houches 1988 proceedings, Session XLIX,
Editors E.~Br\'ezin and J.~Zinn-Justin, North-Holland.})
$$\eqalign{
a&=\sin{\lambda\over 2}(\pi-\theta)\cr
b&=\sin{\lambda\over 2}(\pi+\theta)\cr
c&=\sin\lambda\pi\cr
d&=0\cr
}
$$
where $a$ and $b$ are the weights of two vertices which are
indistinguishable after coupling to gravity (they are related
to each other by a rotation of $\pi/4$); therefore $\theta=0$.
For any value of $0\le\lambda\le 1$,
the model has a continuum limit
which is a conformal field theory with central charge $\cc=1$, and
can be described by a free massless compactified boson, with radius
$R=\sqrt{1-\lambda\over 2}$ (up to some duality
transformations\foot{Here,
we assume that going from the 6-vertex model to the 8-vertex
model correponds to perturbing the free boson
with a {\it magnetic} operator of charge $\pm 4$. The
Kosterlitz--Thouless phase transition occurs at $\lambda=0$, as
it should be.}).

We expect a similar critical behaviour in 
our two matrix model, but modified
by the presence of 2D gravity 
\ref\KPZ{V.~Knizhnik,
A.~Polyakov and A.~Zamolodchikov, {\it Mod. Phys. Lett.} A3 (1988), 819.}. 
This is indeed what we shall find at the critical point of our model,
which occurs for
$\alpha_*=\beta_*=1/(4\pi)$. It corresponds to the 6-vertex model
at the point where
the weight of $(XX^\dagger)^2$ vanishes: $c=0$, or $\lambda=1$.
The 6-vertex model
coupled to gravity has already been studied in
\ref\DAL{S.~Dalley, {\it Mod. Phys. Lett.} A7 (1992), 1651.}
and shown to be equivalent to the standard
formulation of the vortex-free compactified boson coupled to 
gravity \ref\GK{D.~Gross and I.~Klebanov, {\it Nucl. Phys.} B344
(1990), 475.}, with a radius of compactification
$$R={1-\lambda/2\over\sqrt{2}}$$
which is different from the one on the flat square lattice\foot{This is a sign
that the radius of compactification is a function of both
the weights (i.e. of the parameter $\lambda$) {\it and} the
geometry of the lattice. We thank I.~Kostov for pointing
out the discrepancy of radii to us.}. For our critical point,
$\lambda=1$ and we find a radius $R=1/(2\sqrt{2})$, i.e.
one half of the Kosterlitz--Thouless radius. But the $ABAB$ model
is not limited to the 6-vertex model: it
also displays off-critical behaviour
correponding to a particular ``slice'' of the full 8-vertex model.

The plan of the article is as follows: in section 2 we shall
apply the character expansion method to reduce the model to the sum
over highest weights of the $GL(N)$ character expansion. We then derive
the saddle point equations for the density of highest weights and
eigenvalues, which yield
equations defining the characters and the loop averages in the planar
limit.

In section 3 we solve the saddle point equations in terms of
incomplete elliptic integrals and fix the parameters of the solution
by matching it to the large highest weight asymptotics. 

In  section 4 we find the critical line (of the thermodynamical limit
of large lattices) and the critical point of the\rem{$3^{\rm rd}$? 
order} phase transition.
We also describe the critical behaviour around the critical
point, analyse some correlation functions and find the string
susceptibility.

The section 5 is devoted to conclusions.

Finally, the appendices contain a detailed study of three
particular lines in our two-parameter $(\alpha,\beta)$ family
of models; they are the lines given by $\alpha=0$, $\beta=0$,
and $\alpha=\beta$.

\newsec{The model and its representation by a character expansion.}
\subsec{Definition of the model.}
The two-matrix model with $ABAB$ term is given by the partition function
\eqn\defZ{{\rm Z}(\alpha,\beta)=
\int\!\!\!\int dA\, dB\, \exp N\left[-{1\over 2}(\tr A^2+\tr B^2)
+{\alpha\over 4}(\tr A^4+\tr B^4)+{\beta\over 2} \tr(AB)^2\right]}
where $A$ and $B$ are hermitean $N\times N$ matrices. Here $\alpha$
and $\beta$ are positive constants to have positive weights in
the diagrammatic expansion; of course, as usual in matrix models
in connection with quantum gravity, the
analytic continuation from $\alpha,\beta<0$ to
$\alpha,\beta>0$ which allows to define the integral \defZ\ only
makes sense at $N\to\infty$.

As already mentioned in the introduction, there is no obvious
way to do the integration
over the relative ``angle'' $\Omega$ between the two matrices
$A$ and $B$, because no formula is known for the integral
over the unitary group $\int d\Omega
\exp[c \tr (A\Omega B\Omega^\dagger)^2]$. To circumvent this problem,
we use a character expansion of the term $\exp[N{\beta\over2}\tr(AB)^2]$
as a class-function of $AB$: representations of $GL(N)$ are
parametrized by their {\it shifted highest weights}
$h_i=m_i+N-i$ ($i=1\ldots N$), where the $m_i$ are the
standard highest weights. Then one has
$$\eqalign{
{\rm Z}(\alpha,\beta)\sim
\sum_{\{ h\}} &(N\beta/2)^{\# h/2}
{\Delta(h^{\rm even}/2)\over \prod_i (h_i^{\rm even}/2)!}
{\Delta((h^{\rm odd}-1)/2)\over \prod_i ((h_i^{\rm odd}-1)/2)!}\cr
&\int\!\!\!\int dA\, dB\, \exp N\left[-{1\over 2}(\tr A^2+\tr B^2)
+{\alpha\over 4}(\tr A^4+\tr B^4)\right]
\chi_{\{ h\}}(AB)
}$$
where the sum is over all integers $h_i$ that satisfy
$h_1>h_2>\ldots>h_N\ge0$, and
$\# h=\sum m_i=\sum h_i - {N(N-1)\over 2}$ is the
number of boxes of the Young tableau. $\Delta(\cdot)$ is the
Van der Monde determinant, $\chi_{\{ h\}}$ is the $GL(N)$ character
associated to the set of shifted highest weights $\{ h\}$,
and the $h^{\rm even/odd}_i$
are the even/odd $h_i$, which must be in equal numbers. 
It is now possible, using
character orthogonality relations,
to integrate over the relative angle between $A$ and $B$; this
leads to a separation into one-matrix integrals:
\eqn\charexp{
{\rm Z}(\alpha,\beta)\sim \sum_{\{ h\}} (N\beta/2)^{\# h/2} c_{\{h\}} 
[R_{\{ h\}}(\alpha)]^2}
where $c_{\{ h\}}$ is a coefficient:
$$c_{\{ h\}}={1\over\prod_i \lfloor h_i/2\rfloor ! \prod_{i,j}
(h_i^{\rm even}-h_j^{\rm odd})}$$
and $R_{\{ h\}}(\alpha)$ is the one-matrix integral
\eqn\defR{
R_{\{ h\}}(\alpha)=\int dM\, \chi_{\{ h\}}(M) \exp N\left[-{1\over 2}\tr M^2
+{\alpha\over 4}\tr M^4\right]
}
which appears squared
in \charexp\ because the contributions from the two matrices $A$ and
$B$ are identical.

\subsec{Study of $R_{\{ h\}}(\alpha)$}
The one-matrix integral $R_{\{ h\}}$ closely resembles what
was studied in \PZJ\ (see in particular appendix 2), and we
use the same approach. First we go over to eigenvalues:
\eqn\eigenR{
R_{\{ h\}}(\alpha)=\int \prod_k d\lambda_k\, \Delta(\lambda) 
\det\left(\lambda_k^{h_j}\right) \exp N\left[
-{1\over 2}\sum_k \lambda_k^2+{\alpha\over 4}\sum_k \lambda_k^4\right]
}
where $\Delta(\lambda)=\det\left(\lambda_k^{N-j}\right)=
\prod_{j<k}(\lambda_j-\lambda_k)$.

Since we now have $N$ degrees of freedom and an action of order
$N^2$, we can use a saddle point method on the eigenvalues $\lambda_k$.
In order to do so,
we define the resolvent
$$\omega(\lambda)=\atr{1\over \lambda-M}
={1\over N}\sum_k {1\over \lambda-\lambda_k}$$
In the $N\to\infty$ limit, the $\lambda_k$ form a continuous density
on some support $[-\lambda_c,\lambda_c]$, which implies
that $\omega(\lambda)$ has a cut on $[-\lambda_c,\lambda_c]$.
If we introduce the notation 
$\omegasl(\lambda)={1\over2}(\omega(\lambda+i0)+\omega(\lambda-i0))$,
then $\omegasl(\lambda_k)={1\over N}
{\der\over\der\lambda_k}\log\Delta(\lambda)$.

Next we want to differentiate $\det(\lambda_k^{h_j})$;
as $\det(\lambda_k^{h_j})/\Delta(\lambda)$ is a regular function
of the $\lambda_k$ (since for example, it is a Itzykson--Zuber type
integral), we can introduce another function $h(\lambda)$ which
has {\it the same cut} as $\omega(\lambda)$ on $[-\lambda_c,+\lambda_c]$
and such that
\eqn\defh{
\hsl(\lambda_k)={1\over N}\lambda_k {\der\over\der\lambda_k}
\det\left(\lambda_k^{h_j}\right)
}
(we have introduced the factor $\lambda_k$ in front of
the ${\der\over\der\lambda_k}$ for convenience).
The saddle point equation of \defR\ can now be written as
\eqn\spelambda{
\lambda\,\omegasl(\lambda)
+\hsl(\lambda)-\lambda^2+\alpha\lambda^4=0\qquad \lambda
\in [-\lambda_c,\lambda_c]
}
Here comes the key remark: as $\omega(\lambda)$ and $h(\lambda)$ have
the same cut on $[-\lambda_c,\lambda_c$], the slashes can be removed
in \spelambda:
\eqn\spelambdab{
\lambda\,\omega(\lambda)+h^\dagger(\lambda)-\lambda^2+\alpha\lambda^4=0
}
where $h^\dagger$ is $h$ on the other side of the cut $[-\lambda_c,\lambda_c]$.
Expanding in powers of $1/\lambda$ as $\lambda\to\infty$, we find:
\eqn\hexp{
h^\dagger(\lambda)=-\alpha\lambda^4+\lambda^2-1-\sum_{n=1}^\infty
{1\over\lambda^{2n}} \atr{M^{2n}}}

\subsec{Functional inversion and analytic structure of $\lambda(h)$.}
In the same way as we have considered a saddle point on
the eigenvalues, we shall use a saddle point on the highest
weights. Before doing so, we need to understand better
the analytic structure of the functions involved.

In a very
similar fashion as we have defined $\omega(\lambda)$, we define
$H(h)$ to be
$$H(h)=\sum_k {1\over h-h_k}$$
(here, since the $h_k$ scale as $N$ in the large $N$ limit, we do not
need a $1/N$ factor in front). After appropriate rescaling
$h\to h/N$, the $h_k$ tend to a continuous density $\rho(h)$ as
$N\to\infty$. We shall find out that part of the density is
saturated at its maximum value $1$ (the same phenomenon
occurs in e.g. \ref\DK{M.R.~Douglas and V.A.~Kazakov,
{\it Phys. Lett.} B319 (1993), 219.});
therefore we define
the end points $h_1$ and $h_2$ such that
$$\eqalign{
\rho(h)=1\qquad 0<h<h_1\cr
0<\rho(h)<1\qquad h_1<h<h_2\cr}$$
Then $H(h\pm i0)=\Hsl(h)\pm i\pi \rho(h)$ on the cut
$[0,h_2]$, with $\Hsl(h_k)={1\over N} {\der\over\der h_k}\log\Delta(h)$.

Next we introduce the function $L(h)$ which has the same
cut as $H(h)$, and such that
$$\Lsl(h_k)={2\over N} {\der\over\der h_j} \log\det(\lambda_k^{h_j})$$
(the factor of $2$ is due to parity reasons).\rem{this factor
of $2$ is not innocent, it's not just a question of definition!
The point is: with this factor of $2$, $L$ has the same cut as $H$.}
If we define $\lambda(h)=\exp({1\over 2} L(h))$, then,
as proven in a similar context
in the appendix 1 of \PZJ,
$h(\lambda)$ and $\lambda(h)$ are functional inverses of each
other as multi-valued
functions. In particuler, on all sheets of $\lambda(h)$ such that
$\lambda(h)\to\infty$ as $h\to\infty$, one has, according to \hexp,
\eqn\hexpb{
h=-\alpha\lambda^4(h)+\lambda^2(h)-1-\sum_{n=1}^\infty
{1\over\lambda^{2n}(h)} \atr{M^{2n}}}
A similar expansion was found for large $N$ characters in \KaSW\ (see
also \ref\GMKW{D.~Gross and A.~Matytsin,
{\it Nucl. Phys.} B437 (1995), 541\semi
V.A.~Kazakov and T.~Wynter,
{\it Nucl. Phys.} B440 (1995), 407.}).

Inverting the expansion \hexpb\ to express
$\lambda^2$ (we shall from now on
always use $\lambda^2$ and not $\lambda$ for parity reasons) as a
function of $h$, shows that there are two sheets
that satisfy $\lambda^2\to \infty$ as $h\to\infty$: we shall
call these $\lambda^2_\pm$. One of them,
$\lambda_+$ is the ``physical sheet'' i.e.
the original function $\lambda(h)$.
The simplest analytic structure
for $\lambda^2(h)$ is then the following: there is a
semi-infinite cut $[h_3,+\infty]$ connecting $\lambda_+^2$ and
$\lambda_-^2$, the finite cut $[h_1,h_2]$, and a possible
pole/zero at $h=0$. One can then show that $\lambda^2$ satisfies:
\eqn\GG{\lambda_+^2(h) \lambda_-^2(h)= {h\over\alpha}\e{H(h)}}
The figure \cuts\ describes this analytic structure.
\rem{note that here
the factor $h$ creates a double zero instead of cancelling
the pole as for characters. that's why in the end we'll
get incomplete integrals and not complete integrals as in $R^2$ gravity.}
\fig\cuts{Analytic structure of $\lambda^2(h)$. $\lambda^2_+(h)$
has a double zero at $h=0$.}{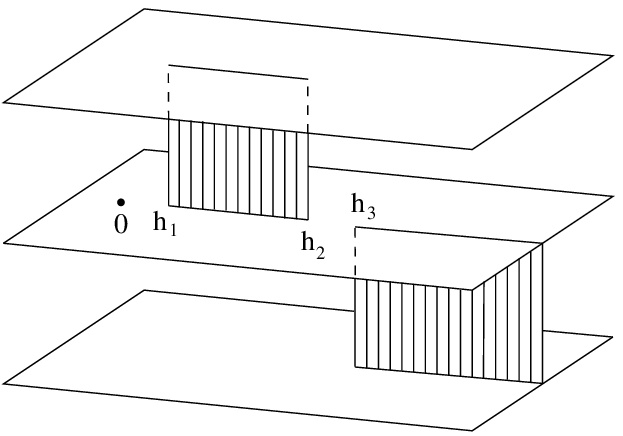}

\subsec{Saddle point equation on the highest weights}
We finally consider the
saddle point equation on $\{ h\}$ in equation \charexp.
From \eigenR\ one infers that 
$2(d/dh) \log R_h\equiv\Lsl(h)$\rem{in old notation $L(h)=\log G(h)$},
while $c_{\{ h\}}$ has the following large $N$ limit:
$$\log c_{\{ h\}}=-\sum_i {h_i\over 2}\left(\log{h_i\over 2}-1\right)
-{1\over 2}\log\Delta(h)$$
so that, after rescaling of the $h_i$,
$(d/dh) \log (c_{\{ h\}}/N^{\# h/2})=-(\Hsl(h)+\log {h\over 2})/2$. Therefore
the saddle point equation reads:
\eqn\spe{\Lsl(h) - {\Hsl(h)\over 2} = {1\over 2} \log(h/\beta)
\quad\forall h\in[h_1,h_2]}

\newsec{Solution of the saddle point equations in terms of elliptic functions.}
In order to solve the saddle point equations, we need to 
remove the logarithmic cuts of the function $2L(h)-H(h)$ on
the physical sheet.
This is achieved by defining the function
$D(h)=2L(h)-H(h)-3\log h +\log(h-h_1)$.
$D$ has only two (square root type) cuts: $[h_1,h_2]$ and
$[h_3,+\infty]$, and it satisfies the following equations:
\eqn\twospe{\eqalign{
\Dsl&=\log{h-h_1\over\beta h^2}\quad\forall h\in [h_1,h_2]\cr
\Dsl&=\log{h-h_1\over\alpha  h^2}\quad\forall h\in [h_3,+\infty]\cr
}}

The solution of \twospe\ is given in terms of elliptic functions
(for some useful formulae on elliptic functions,
see \ref\LAWD{D.F.~Lawden, ``Elliptic Functions and Applications'',
Springer, New York, 1989.}).
Define $r(h)=\sqrt{(h-h_1)(h-h_2)(h-h_3)}$ and the incomplete
elliptic integral of the third kind
($\tilde{h}$ real):
\eqn\defell{\Phi_{\tilde{h}}(h)=r(h) \int_{\tilde{h}+i0}^{+\infty}
{dh'\over (h-h') r(h')}}
Then $D(h)$ is given by
\eqn\solD{D(h)=\log{h-h_1\over\beta h^2} 
-\Phi_{h_3}(h){\log(\beta/\alpha)\over i\pi}
-\Phi_{h_1}(h)+2\Phi_0(h)}

Next define elliptic parametrizations of $h$: $x$ is such that
$$\sn(x,k)\equiv\sn x=\sqrt{h_3-h_1\over h_3-h}$$
where $k=\sqrt{h_3-h_2\over h_3-h_1}$. $x$ is chosen such that
$0\le \Re x\le K$, $-iK'\le \Im x\le iK'$
(where $K$ and $K'$ are the quarter-periods), and $x(h)$
has a cut on $[h_1,+\infty]$.
Also define the rescaled $\Theta$ functions:
$\Theta_a(x)=\theta_a(\pi x/2K)$, where the $\theta_a$, $a=1\ldots 4$,
are the usual $\theta$ functions. Finally, let $Z_a(x)=(d/dx)
\log\Theta_a(x)$.

Then $\Phi_{\tilde{h}}(h)$ is given explicitly by:
\eqn\expell{\Phi_{\tilde{h}}(h)=\tilde{x} \left[ 2Z_1(x)-{i\pi\over K}\right]
+\log{\Theta_1(x-\tilde{x})\over\Theta_1(x+\tilde{x})}}
where $\tilde{x}=x(h=\tilde{h})$.

Next we must adjust the behaviour of $D(h)$ as $h\to\infty$,
which will fix the unknown constants $h_1$, $h_2$, $h_3$.
Starting from
$h=-\alpha \lambda^4 + \lambda^2-1+O(1/\lambda^2)$
(Eq. \hexpb), one proves that
\eqn\asyG{2 L_\pm(h)
=\log{-h\over\alpha} \mp {1\over\sqrt{-\alpha h}} + {1\over h}
+ O(h^{-3/2})}
\rem{how to distinguish $L_+$ from $L_-$?}with $L\equiv L_+$.
As $h\to\infty$, $x\sim \sqrt{h_3-h_1}/\sqrt{-h}$;
when $x\to 0$, $Z_1(x)\sim 1/x$, so there is a divergent part in \solD\ 
which must cancel out; this leads to the first condition:
($x(h=0)\equiv x_0$, 
$x(h=h_1)=K$, $x(h=h_2+i0)=K+i K'$, $x(h=h_3+i0)=iK'$)
\eqn\condone{\llap{condition \# 1:\hskip2.2cm}
2x_0=K+{\log(\beta/\alpha)\over \pi} K'}
Using \condone\ and the explicit
expressions of $\Phi_{h_1}$ and $\Phi_{h_3}$, 
$D(h)$ can be rewritten in the simpler form
\eqn\solDb{D(h)=\log{h-h_1\over -\alpha h^2}
- {\log(\beta/\alpha)\over K} x(h) 
+ 2 \log{\Theta_1(x_0-x(h))\over\Theta_1(x_0+x(h))}}
or alternatively as a function of $y\equiv x-K$:
\eqn\solDc{D(h)=\log{h-h_1\over -\alpha h^2} 
- {\log(\beta/\alpha)\over K} y(h) 
+ 2 \log{\Theta_2(x_0-y(h))\over\Theta_2(x_0+y(h))}}

By matching the $1/\sqrt{h}$ terms in the $h\to\infty$ asymptotics,
one obtains the second condition:
\eqn\condtwo{\llap{condition \# 2:\hskip2.4cm} \Omega_1
={1\over\sqrt{\alpha(h_3-h_1)}}}
We have defined: 
$$\Omega_a\equiv {\log(\beta/\alpha)\over K}+4Z_a(x_0)\qquad a=1\ldots4$$

From \solDc\ we obtain the expression for the density of 
highest weights:
\eqn\DEN{\rho(h)=1+{\log(\beta/\alpha)\over i\pi K} y(h)
-{2\over i\pi} \log{\Theta_2(x_0-y(h))\over\Theta_2(x_0+y(h))}}
When $h\in [h_1,h_2]$, $y$ is purely imaginary,
$\Im y\in[0,K']$, so that $\rho(h)$ is real.
\fig\dens{Density $\rho(h)$, $q=0.1$,
$\alpha=\beta=0.062674\ldots$.}{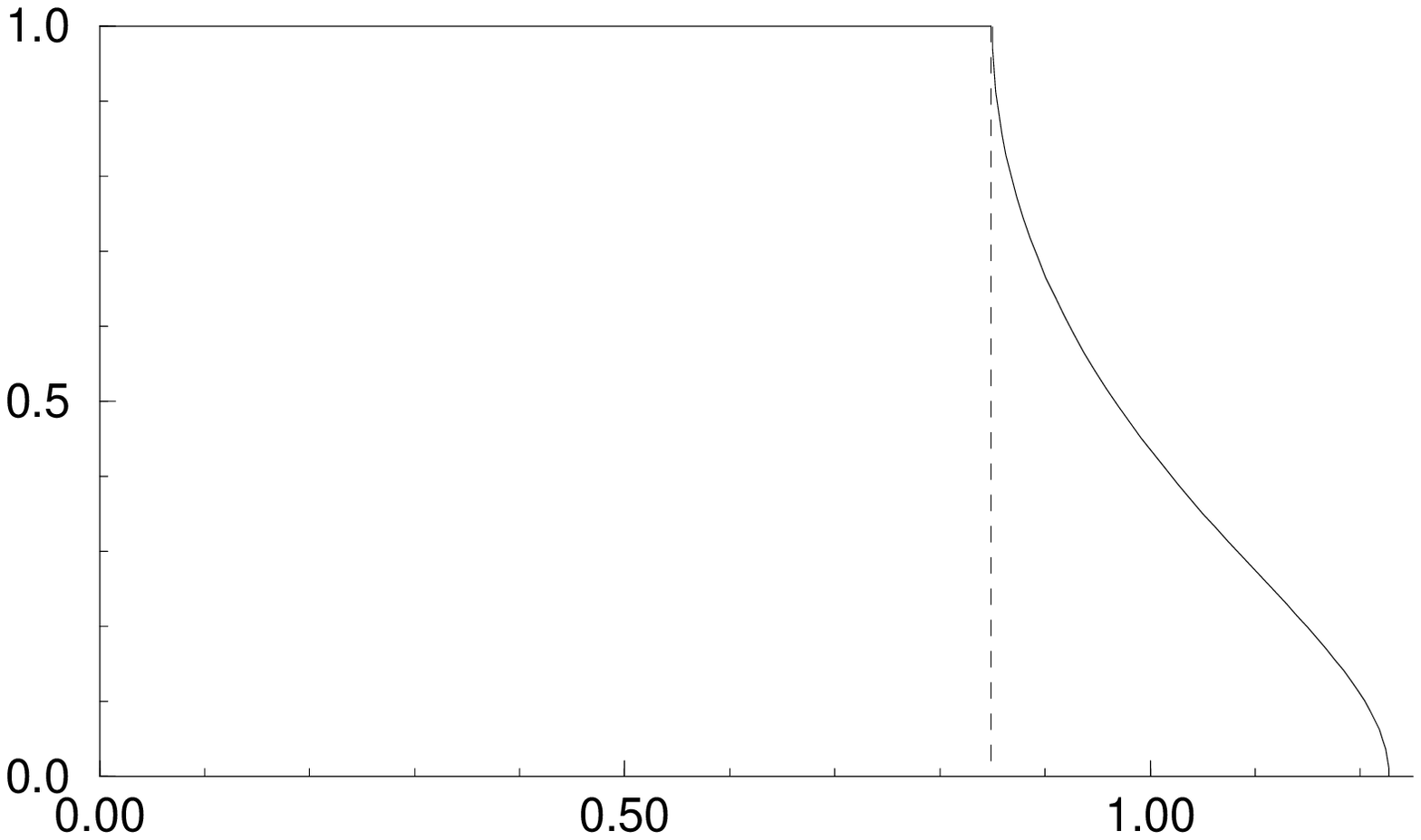}
To fix $h_1$, $h_2$ and $h_3$, we need a third condition, which
is the normalisation of the density $\rho(h)$:
$$\int_{h_1}^{h_2} dh\, \rho(h)=1-h_1$$
Integrating by parts and changing variables from $h$ to $y$ yields:
\eqn\normden{\int_0^{iK'} dy\, h(y)\left[{\log(\beta/\alpha)\over K}
-2 (Z_2(y-x_0)-Z_2(y+x_0))\right]=-i\pi}
As $Z_2(y-x_0)-
Z_2(y+x_0)$ is an elliptic function:
$${1\over 2}\left[Z_2(y-x_0)-Z_2(y+x_0)\right]=-Z_4(x_0)+
\cn x_0 \dn x_0 \sn x_0
{\dc^2 y\over 1-\sn^2 x_0 \dc^2 y}
$$
($\dc\equiv {\displaystyle \dn\over\displaystyle \cn}$).
one can express \normden\ in terms of the
complete elliptic integrals of the first and second kinds. The
resulting equation fixes $\alpha$:
\eqn\condthree{\llap{condition \# 3:\hskip1.5cm}
\alpha={1\over\pi\Omega_1^{\,\,2}}
\left(E'\Omega_1-K'\Omega_4 {1\over \sn^2 x_0}\right)}
To summarize, given $\log(\beta/\alpha)$ and 
the elliptic nome $q=\e{-\pi K'/K}$,
one can compute $x_0$ from \condone, then $\alpha$ and $\beta$
from \condthree, and finally $h_1$, $h_2$, $h_3$ from \condtwo.

\newsec{The critical line and the critical point.}
We now investigate the critical properties of the model.
Just as in the standard eigenvalue problem,
criticality is attained when a branch point
collides with a critical point (i.e. a point where the derivative
vanishes), so that the square root singularity
degenerates into a $(h-h_0)^{3/2}$ behavior.\rem{note that
the reason it is the same trick as in the standard eigenvalue
problem is that this behavior is ``self-dual'' under functional
inversion.}
If one starts with small coupling constants
$\alpha$ and $\beta$ and then increases these constants, two
phenomena can occur, which correpond to two pieces
of the critical line.
\subsec{Criticality of type A.}
The first type of criticality appears when the singularity
appears in the vicinity of $h_3$, i.e. at the start
of the infinite cut $[h_3,+\infty]$. This is what
happens, for example, if one fixes $\beta$ to a small value,
and increases $\alpha$: for $\alpha$ smaller than
a critical value $\alpha_c(\beta)$, $h_3$ is well-defined and real,
but for $\alpha>\alpha_c$ $h_3$ becomes complex, which is the sign
of a change of analytic structure, and therefore of criticality.

In order to find the critical line in the $(\alpha,\beta)$ plane,
one looks at the behavior of $D(h)$ as $h\to
h_3$, i.e. $x\to \pm iK'$. The procedure is very similar to looking
at the $x\to 0$ behaviour; criticality is obtained by cancelling
the $\sqrt{h-h_3}$ term, that is when
\eqn\critone{\Omega_4={\log(\beta/\alpha)\over K} + 4 Z_4(x_0)=0}
(compare with \condtwo).

Using \condtwo\ one can simplify this condition:
$${\cn x_0 \dn x_0\over \sn x_0} = {1\over
4\sqrt{\alpha(h_3-h_1)}}$$
and reexpress it in terms of $h$:
\eqn\critoneb{{h_1 h_2\over h_3}={1\over 16\alpha}}

\subsec{Criticality of type B.}
This time the singularity occurs at the level of the
finite cut $[h_1,h_2]$; this is what happens for small $\alpha$
and increasing $\beta$.

Criticality is more precisely found by looking at the behaviour of $D(h)$ as
$h\to h_2$, i.e. $x\to K\pm iK'$. Cancellation
of the $\sqrt{h-h_2}$ term implies:
\eqn\crittwo{\Omega_3={\log(\beta/\alpha)\over K} + 4 Z_3(x_0)=0}
Combining it with \condtwo\ yields
$${\cn x_0\over \sn x_0 \dn x_0} = {1\over
4\sqrt{\alpha(h_3-h_1)}}$$
or equivalently
\eqn\crittwob{{h_1 h_3\over h_2}={1\over 16\alpha}}

\subsec{Double criticality: the critical point.}
When one imposes both \critoneb\ and \crittwob,
one finds $h_3=h_2$, $h_1=1/(16\alpha)$: the critical point
correponds to the limit when the two cuts touch each other.
Therefore we are
in the trigonometric limit $k\to 0$. We also find $\alpha=\beta$.
In this case one can solve directly the equations \twospe\ which combine
into a single equation; the result is:
$$D(h)=\log{h-h_1\over -\alpha h^2} -2 \log{\sqrt{h_1-h}+\sqrt{h_1}
\over -\sqrt{h_1-h}+\sqrt{h_1}}=\log{h-h_1\over -\alpha}
-4\log(\sqrt{h_1-h}+\sqrt{h_1})$$
$$\rho(h)=1-{4\over \pi}
\arctan\sqrt{h/h_1-1}$$
$\rho(h)$ is zero for $h=2h_1$, so one has $h_2=h_3=2h_1$.
Finally the normalisation of the density gives
$$h_1={\pi\over 4}$$
and therefore $\alpha=\beta=1/(4\pi)$.

The critical line and the critical point have been drawn numerically
on figure \curve. The critical line is divided in two: the left of
the critical point ($\alpha>\beta$) is the phase A (criticality
is of type A),
whereas the right ($\beta>\alpha$) is the phase B.
Note that the trigonometric limit $q\to 0$
is precisely the neighborhood of the critical point, and on the
contrary that the hyperbolic limit $q\to 1$ is the
limit $\alpha\to 0$ or $\beta\to 0$.

\fig\curve{Critical line, critical point
and equipotentials of $q$ ($q=0(0.03)0.6$) 
in the $(\alpha,\beta)$ plane.}{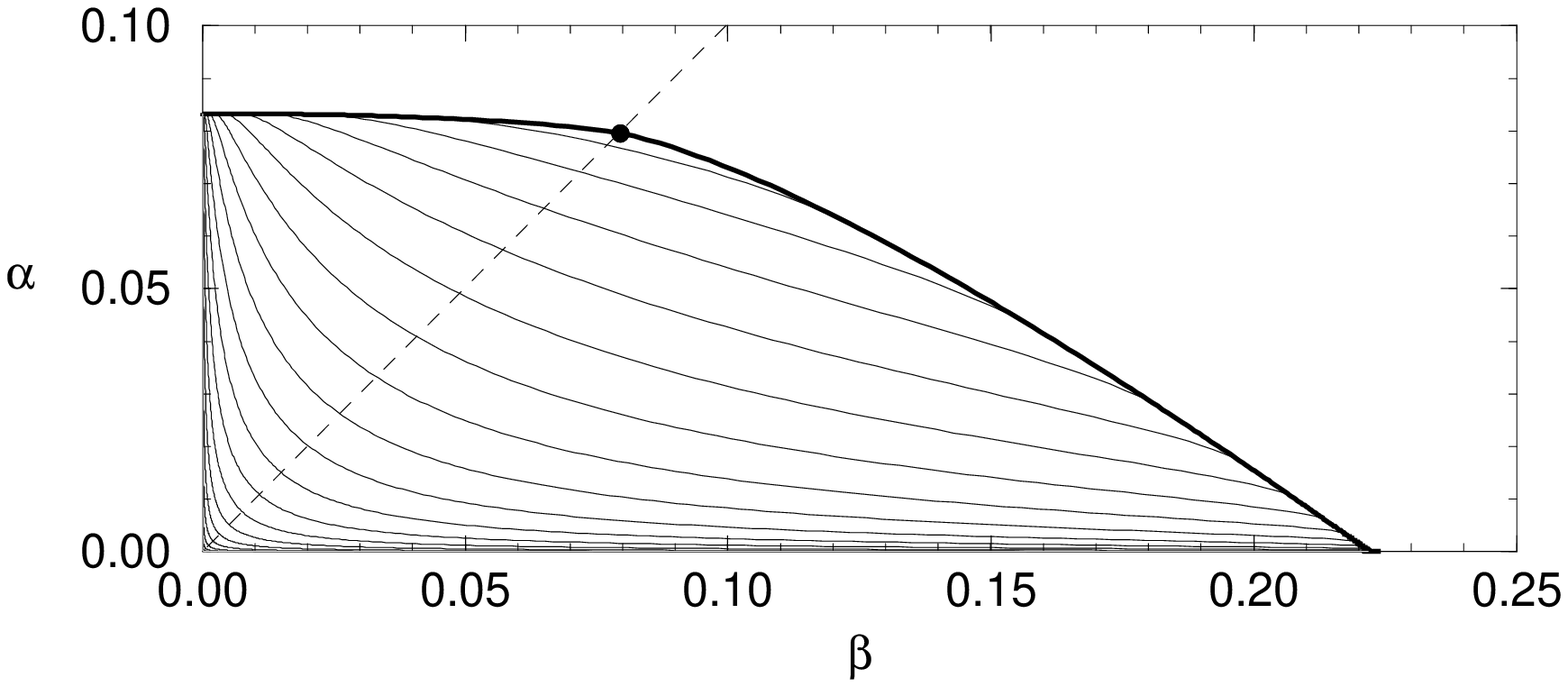}

\fig\curveb{$h_1$, $h_2$, $h_3$ {\it on the critical line}
as a function of $\beta$.}{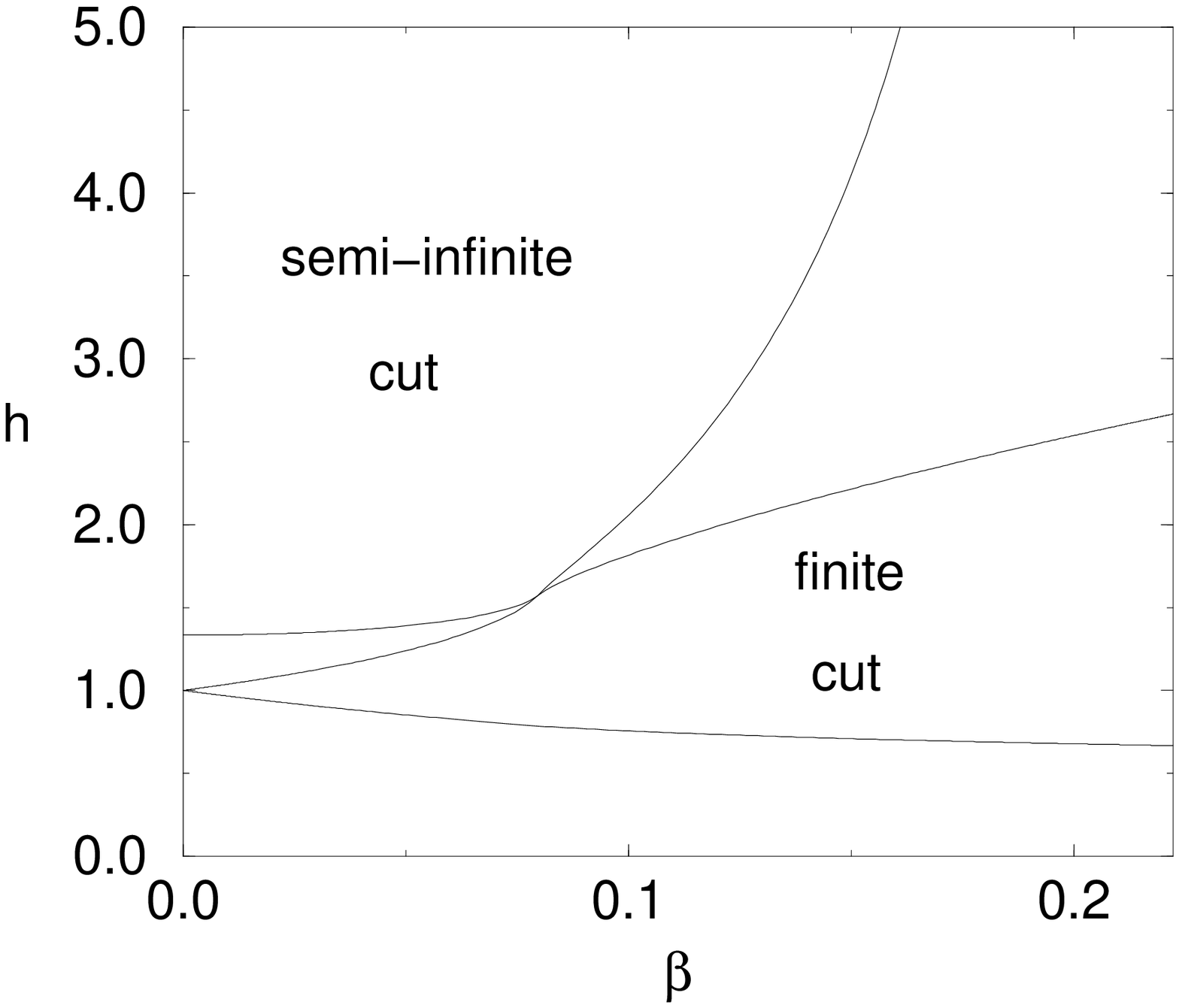}

To complete the picture {\it at the critical point}, we need to go back from $D(h)$ to
$L(h)$, which requires to compute $H(h)$. In fact, $H(h)$ is not a 
standard function but $dH/dh$ is:
$$\eqalign{
{dH\over dh}&=\int_0^{h_2} \rho(h'){dh'\over (h-h')^2}\cr
&=-{2\over\pi} {1\over h} \sqrt{h_1\over h-h_1}
\log{\sqrt{h-h_1}+\sqrt{h_1}\over\sqrt{h-h_1}-\sqrt{h_1}}\cr}$$
This expression is sufficient to expand $2L(h)=D(h)-H(h)+3\log
h-\log(h-h_1)$ around the ``critical point'' $h_c\equiv h_2=h_3$,
which provides us with the loop function $h(\lambda)$ for large loops.
Only $H(h)$ has a singularity at this point:
$dH/dh \propto \log(h-h_c)$,
and therefore $H(h)\propto (h-h_c)\log(h-h_c)$, or finally
$$
\lambda-\lambda_c \propto (h-h_c)\log(h-h_c)
$$
which we can invert:
\eqn\loopcrit{
h-h_c \propto {\lambda-\lambda_c\over\log(\lambda-\lambda_c)}
}
\rem{comment?}
We shall now study the properties of the vicinity of the critical
line.\rem{etc}
In particular, we will need the exact expression for
a correlation function. By functional
inversion, $\lambda(h)$ provides us with all correlation functions
of the type $\left<\tr A^{2n}\right>$. In practice,
one can easily compute $\left<\tr A^2\right>$ by
inverting \hexpb\ up to order $1/\lambda^2$:
\eqn\asyGb{2 L(h)
=\log{-h\over\alpha} - {1\over\sqrt{-\alpha h}} + {1\over h}
+ \left(\sqrt{\alpha}\atr{A^2}+{1\over
24\alpha^{3/2}}-{1\over 2\sqrt{\alpha}}
\right){1\over 2 (-h)^{3/2}}+O(h^{-2})}
where $L(h)=2\log\lambda(h)$, and expanding the explicit
expression \solDb\ of $D(h)$.

Here, we shall use another correlation function, $\left<\tr(AB)^2\right>$. It is given by:
\eqn\ABAB{\eqalign{
\atr{(AB)^2}&={\der\over\der\beta} Z\cr
&={1\over\beta} \left< \# h \right>\cr
&={1\over\beta} \left[ {1\over 2} h_1^2 -{1\over 2} 
+ \int_{h_1}^{h_2} \rho(h) h\, dh\right]\cr
&={1\over\beta}\left[-{1\over 2}-\int_0^{iK'} dy {d\rho\over dy} {h^2\over 2}
\right]\cr}
}
At this point, the calculation becomes similar to that of $\alpha$.
The result is:
\eqn\ABABb{
\atr{(AB)^2}=
{1\over\beta}\left[-{1\over 2}+h_3+\pi
{E'(\Omega_4/\sn^2 x_0 - \Omega_1 {2\over 3} (2-k'^2))+K'\Omega_1 {1\over 3} (1-k'^2)
\over (E'\Omega_1-K'\Omega_4/\sn^2 x_0)^2}\right]
}

\subsec{Vicinity of the critical line.}
Let us suppose first that we approach the critical line at fixed slope
$s\ne 0$: then $q$ tends to a non-zero critical value $q_c$, and
one can show the following facts: ${d\over d q} \alpha_{|q=q_c}=0$,
${d\over d q} \left<\tr (AB)^2\right>_{q=q_c}=0$ and also
${d\over d q} h_k{}_{|q=q_c}=0$ with $k=1,2$ for criticality
of type A, $k=1,3$ for criticality of type B.
This implies in particular that
$$\atr{(AB)^2}_{\rm sing} \sim \Delta^{3/2}$$
where $\Delta\equiv \alpha_c-\alpha$ is the renormalized cosmological
constant,
and therefore the string susceptibility exponent is $-{1\over 2}$:
this means that on the whole critical line except at the critical
point, the continuum theory is a $\cc=0$ theory (pure gravity).

In fact, the analytic structure found around any point on the
critical line in phase A (resp. B) is identical
to that found for the endpoint $\alpha=1/12$, $\beta=0$ (resp.
$\beta=2/9$, $\alpha=0$). These two particular points have been
studied in detail in appendix A. The conclusion one draws from
the more explicit calculations done at these points is the following:
though we have in both phases a $\cc=0$ theory, for phase A the
loop scaling function is the usual pure gravity loop function, whereas
in phase B this function describes non-trivial loops
(eq. \ntloop).

\subsec{Vicinity of the critical point.}
We now investigate the
vicinity of the critical point $\alpha=\beta=1/(4\pi)$, 
which is also the small $q$ region.
We refer to appendix B for details on small $q$ expansions.

If we define the deviation from criticality
$$\Delta\equiv 1-4\pi \alpha$$
and the slope in the $(\alpha,\beta)$ plane
$$s\equiv{\log(\beta/\alpha)\over 4\pi}$$
then we obtain up to order $q^3$ 
the equation defining $q$ in terms of $\alpha$ and $\beta$:
\eqn\EQST{
\Delta=2s+8q^2\log q^{-1} +2s^2 \log^2 q^{-1}+4q^2 -4s^2
}

We can find from here the shape of the critical line near the critical point.
It is given by the equation:
\eqn\CRLI{{d\over dq} \Delta = 16q\log q^{-1} -{4s^2\over q} \log q^{-1}=0}
or 
\eqn\CRLI{
q= \pm s/2
}
for $\beta <\alpha$ and $\beta >\alpha$, respectively. 
One can check (using the last two expansions of \VAREX) that it fits well the conditions 
\critone\ and \crittwo. 

In particular, this implies that the slope of this line at the
critical point is:
\eqn\SLOPE{
{d\alpha \over d\beta}={1\over 1- 2\pi}
}
\rem{It is left to define from \EQST\ the scaling function $q({\Delta
\over s^\alpha} \log\cdots)$...}

It is possible to expand \ABABb\ around $q=0$; the result is of the form
\eqn\ABABc{
\atr{(AB)^2}=
c_1+c_2 s+c_3 s^2 \log^2 q^{-1}
+c_4 s^2\log q^{-1} +c_5 q^2 \log q^{-1}+c_6 s^2+c_7 q^2
}
where the coefficients $c_i$ are given in appendix B.

For $s=0$, from \EQST\ $\Delta\propto q^2\log q^{-1}$, and the singular part
of \ABABc\ yields
\eqn\ABABsing{\atr{(AB)^2}_{\rm sing}
\propto {\Delta\over \log\Delta}}
that is a zero string susceptibility exponent, plus
logarithmic corrections.
This is characteristic of a $\cc=1$ model coupled to gravity
\ref\KAZMIG{V.A.~Kazakov and A.~Migdal, {\it Nucl. Phys.} B311 (1988), 171.}.

Let us also note that on the critical line $s=\pm 2q$, one can check that,
as expected,
$${d\over dq}\atr{(AB)^2}=0$$

The free energy of the underlying statistical-mechanical model in the
thermodynamical limit can be obtained by the following trick \KAZBOUL.
The planar free energy of the matrix model 
(i.e. the partition function of the statistical model on connected
planar graphs) near the critical line looks as
\eqn\CRIF{
\log {\rm Z} \sim \sum_n n^{\gamma_{str}-3} [\alpha/\alpha_*(s)]^n
}
where $\alpha_*(s)$ is the radius of convergency of the series at
fixed $s$, and $n$ (the number of $A^4$ vertices) characterizes
the size of the graphs.
Therefore, 
$\log[\alpha_*(s)]$ plays the role of the free energy per volume unit
in the thermodynamical limit. By plugging \CRLI\ into
\EQST\ we find:
\eqn\DELCR{
\Delta_*(s)=1-4\pi \alpha_*(s)=2s+s^2\log(4/s^2)+s^2\log^2(4/s^2)-3s^2
}
 
Whereas this function and its first derivative are finite and
continuous the second derivative (playing the role of the specific heat
of the problem with $s$ along the critical line being the
``temperature'')
developes a $\log^2$ type singularity:
\eqn\LOGS{
{d^2 \over ds^2} \Delta_*(s) \sim  2\log^2(4/s^2) \qquad \hbox{for $s
\rightarrow 0$}
}
giving rise to a second order
phase transition. 

This is an unusual result since for most of the spin
models on random dynamical graphs the transition is usually of the
third order. A similar situation, but with a $\log$ type
singularity accurs in the model of percolation on dynamical random
graphs solved in \ref\KAZPER{V.A.~Kazakov, {\it Mod. Phys. Lett.} A4 (1989),
1691.}.

\newsec{Conclusion}
Our results show that the possibilities to find new exact solutions of
the matrix models are far from being exhausted. The character
expansion method used here enlarges the class of such models with a
potentially very rich spectrum of critical behaviours and thus with
numerous possible applications to
mesoscopic physics, enumeration of graphs, statistical-mechanical
models on random dynamical graphs, etc. 

For example, a rather general
class of such models for $p$ hermitean matrices can be given by the action:
$S=\tr(\sum_{k=1}^p V_k(M_k) + f(\prod_{k=1}^p M_k))$ for  arbitrary
functions $V_k$ and $F$. It is easy to see from the orthogonality of
characters that the exponent of the last term can be integrated over
relative angles of the matrices. But even this is not the most general
example.

One interesting question we can pose here is the following:
is there any integrable
structure behind the representation expansions of partition functions
like \charexp? For more conventional matrix models, we are used to the
fact that they can be represented as tau-functiopns of some integrable
hierarchies of classical differential equations. These tau-functions
are usually some $N\times N$ determinants or Fredholm determinants in
the grand canonical ensemble with respect to $N$. 

The particular elements of this formula remind the objects
known from the tau-functions: the characters or their integrals with
one-matrix weights (like in \defR), as well as the product
$\prod(h^{\rm even}-h^{\rm odd})^{-1}$ have
determinant representations (or, in other
words, some fermionic analogues). But we don't know such a
representation for the whole sum over Young tableaux. If such a
representation  exists it could allow us to analyse the critical
behaviour for all genera of random graphs in the double scaling
limit, like for the more standard matrix models
\ref\KB{E.~Br\'ezin and V.~Kazakov, {\it Phys. Lett.} B236 (1990), 144.},
\ref\DS{M.~Douglas and S.~Shenker, {\it Nucl. Phys.} B335 (1990), 635.},
\ref\GM{D.~Gross and A.~Migdal, {\it Phys. Rev. Lett.} 64 (1990), 127.}.

Let us also note that the universal behaviours of the
physical quantities we found,
like the loop function \loopcrit\ look
different from the known patterns for the $\cc=1$ matter
coupled to 2D gravity. A possible explanation 
is that the loops we are considering
differ from the usual ones (through
different ``boundary conditions'' on the loop). 
If so, it would be interesting to make the connection between
the different types of loops.

Also, the
critical point we have in our model is only
one point in the critical line of $\cc=1$ CFTs.
In order to describe more precisely the properties of this point,
it would be interesting to investigate the possibility of
a non-trivial scaling limit around
the critical point $\alpha_\star=\beta_\star=1/(4\pi)$.
This would shed light on
the perturbing operator around the fixed point, which is directly
related to the radius of compactification
$R$ in the equivalent bosonic
formulation; it would give an independent check of the predicted radius
$R=1/(2\sqrt{2})$.

Finally, it would of course be very interesting to obtain the
solution of the full 8-vertex model coupled to 2D gravity (with all
couplings $b$, $c$, $d$ independent
in \onecmm). This would allow to describe the
most general critical properties of $\cc=1$ coupled to 2D gravity.
But for the moment, even character
expansion methods cannot solve this more general
matrix model.
\bigbreak\bigskip
\centerline{\bf Acknowledgements}\nobreak
The authors would like to thank I.~Kostov for many useful
discussions.
\vfill\eject
\appendix{A}{Study of particular cases.}
\subsec{$\beta=0$: the usual one matrix model revisited.}
When $\beta=0$ the model reduces to two decoupled one-matrix models
with the standard action ${1\over2}\tr M^2+{\alpha\over4}\tr M^4$.
We shall now compare our solution with the classical solution
of pure gravity.

Since the expansion in characters is trivial for $\beta=0$,
the saddle point character is necessarily the trivial representation,
i.e. $\rho(h)=1$ for $h\in [0,1]$. Therefore
$$H(h)=\log {h\over h-1}$$
which implies that  $h_1=h_2=1$ (the cut reduces to a pole at
$h=1$), and
\eqn\Gpur{
\lambda^2_\pm(h)={1\over \sqrt{\alpha}} 
{(\sqrt{h_3}\mp \sqrt{h_3-h})^2
\over\sqrt{h_3-1}\mp \sqrt{h_3-h}}
}
$h_3$ is given explicitly as a function of $\alpha$ by
$$h_3={5/4-3\alpha+\sqrt{1-12\alpha}\over 9\alpha}$$
which has a singularity at $\alpha=1/12$, the critical point
of pure gravity.

On the other hand, one can use the standard
method for this model, which is to solve directly
the saddle point equations for the eigenvalues.
Indeed, when the character is trivial, it
is clear from \defh\ that $h(\lambda)=\lambda \omega(\lambda)$;
then \spelambda\ becomes
$$2\hsl-\lambda^2+\alpha\lambda^4=0$$
which yields
\eqn\hpur{
h_\pm={1\over 2}\left[\lambda^2-\alpha\lambda^2
\pm(\alpha\lambda^2-1+{1\over 2}\alpha
\lambda_c^2)\sqrt{\lambda^2(\lambda^2-\lambda_c^2)}\right]
}
where $h(\lambda)\equiv h_+(\lambda)$ is the physical sheet,
and $h^\dagger(\lambda)\equiv h_-(\lambda)$ is the other sheet connected
by the cut $[-\lambda_c,+\lambda_c]$\foot{As a function of
$\lambda^2$, the cut of $h(\lambda^2)$ is $[0,\lambda_c^2]$.}.
The end point of the cut $\lambda_c$ is given by:
$$\lambda_c^2={2\over 3\alpha}(1-\sqrt{1-12\alpha})$$
Note that $h^\dagger(\lambda)$ has the
correct expansion \hexp\ as $\lambda\to\infty$.

It is easy to show that $\lambda^2(h)$ given by \Gpur, and
$h(\lambda^2)$ given by \hpur, are functional inverses of each
other: indeed they are the solution of the equation
\eqn\quadequ{
\alpha \lambda^4(h-1)-\lambda^2(h-\alpha h_3 \lambda_c^2)+h^2=0
}
which is quadratic in $\lambda^2$ and in $h$.
\rem{diagrammatic proof? (loop equation?)}

Criticality is
of course obtained when $\alpha\to\alpha_c=1/12$. By computing correlation
functions one immediately finds that the string
susceptiblity $\gamma=-{1\over 2}$, that is pure gravity.

Let us also remind the reader of the universal loop scaling function.
The asymptotics of the loop average
$\left<\tr A^{2n}\right>$ (which is interpreted
as the summation over surfaces with a fixed boundary of length $n$),
$n$ large, are dominated by the singularity of the
resolvent $\omega(\lambda)$ (or of $h(\lambda)$)
which is closest to $\lambda=\infty$.
Here, it is the square root singularity starting at
$\lambda=\lambda_c$. Equivalently, one can say that
$h_c\equiv h(\lambda=\lambda_c)$
is a critical point of the function $\lambda(h)$. When
$\alpha\to\alpha_c$, $h_c$ collides with the branch point $h_3$,
as expected.
One can now take a scaling limit where the renormalized
cosmological constant $\Delta=\alpha_c-\alpha\to 0$ (so that the
average area of the closed surface
$\left<{\cal A}\right>\propto 1/\Delta$ diverges)
and the renormalized boundary cosmological constant
$\lambda-\lambda_c\to 0$
(so that the typical size of the loop $\left<l\right>\propto
1/(\lambda-\lambda_c)$ diverges) in a correlated manner.
Here, the scaling function is:
\eqn\tloop{
{h-h_c\over\Delta^{3/4}}\propto 
\left({\rm cst}-{\lambda-\lambda_c\over\Delta^{1/2}}\right)
\left({\lambda-\lambda_c\over\Delta^{1/2}}\right)^{1/2}
}

\subsec{$\alpha=0$.}
When $\alpha=0$, only the vertex $\tr(ABAB)$ survives,
so that one has a model with two types of loops which
intersect each other all over the surface\rem{figure?}.
Discarding one of the two types of loops, one finds
a model of {\it dense} loops, i.e. loops that cover
the entire space. This looks very similar to
the dense phase of the $O(1)$
model (i.e. low-temperature phase of the Ising model
coupled to gravity), which has already been studied in \ref\KoS{
I.~Kostov, {\it Mod. Phys. Lett.} A4 (1989), 217\semi
I.~Kostov and M.~Staudacher, {\it Nucl. Phys.} B384 (1992), 459.}. 
In fact, the $ABAB$ model with $\alpha=0$ (and its $O(n)$ generalisation)
was already studied in \ref\ChCh{L.~Chekhov and
C.~Kristjansen, {\it Nucl. Phys.} B479 (1996), 683.}
(by completely different methods from ours),
where it was shown to be equivalent to a modified
$O(1)$ model. Let us compare here our results with those of \KoS\ and \ChCh.

If $\alpha=0$, the expansion \hexpb\ implies that
the analytic structure looks different from the generic case
$\alpha>0$: indeed, $\lambda$ has no cut going to infinity.
In other words, when $\alpha\to 0$, $h_3$ tends to infinity
(more precisely, one can show that $h_3\sim 1/(4\alpha)$).
We are left with a single cut $[h_1,h_2]$;
using the general solution and the additional relation
$\lambda^2(h)=h \exp(H(h))$ which follows from the fact that
$\alpha=0$, we can explicitly the function $\lambda(h)$:
\eqn\alphzero{
\lambda(h)={1\over 2}{(h+\sqrt{(h-h_1)(h-h_2)}+\sqrt{h_1 h_2})^2
\over h(\sqrt{h-h_1}+\sqrt{h-h_2})}
}
$h_1$ and $h_2$ are given as solutions of a third degree
equation: $h_{1,2}=\pm {\beta\over 8} x^2 + {1\over 3} x -{1\over 3}$
where
$$ x^3-{16\over 3\beta^2}x+{64\over 3\beta^2}=0$$
The discriminant $\Delta'\equiv {4\over 81\beta^2}-1$ vanishes
when $\beta=\beta_c={2\over 9}$ (the value of $\beta_c$ coincides
with what was found in \ChCh). Then $h_1={2\over 3}$,
$h_2={8\over 3}$, and $\lambda_c=\lambda(h=h_2)=3/\sqrt{2}$
satisfies $\beta_c \lambda_c^2=1$, as follows from
general arguments \ChCh.

One can easily compute the string susceptibility in the limit
$\beta\to\beta_c$; using
$x=6(1+{1\over\sqrt{3}} \Delta'{}^{1/2}+\cdots)$, one finds
$$\atr{A^2}_{\rm sing}\propto \Delta'{}^{3/2}$$
so that the string susceptibility exponent $\gamma=-1/2$, which indicates
a $\cc=0$ model.

However, the loop scaling function is non-trivial; in order
to see this, we use the simple formula:
$${d\over dh} L(h)=2\lambda^{-1}{d\lambda\over dh}
={h-2\sqrt{h_1 h_2}\over h\sqrt{(h-h_1)(h-h_2)}}$$
which shows that $\lambda(h)$ has a critical point
at $h_c=2\sqrt{h_1 h_2}$. This point meets the branch point $h_2$
when $\beta\to\beta_c$, as expected.
In the limit $\beta\to\beta_c$, the proper scaling ansatz
is that $h-h_c\propto \Delta'{}^{1/2}$, 
$\lambda-\lambda_c\propto \Delta'{}^{3/4}$.
Then the universal loop scaling function is given by:
\eqn\ntloop{
{\lambda-\lambda_c\over\Delta'{}^{3/4}}\propto
{\left({\displaystyle h-h_c\over\displaystyle\Delta'{}^{1/2}}\right)^2
\over \left({\rm cst}+{\displaystyle h-h_c\over\displaystyle
\Delta'{}^{1/2}}\right)^{1/2}}
}
Note that the scaling $\lambda-\lambda_c\propto\Delta'{}^{3/4}$
is unusual since it implies that $\left<l\right>\propto
\left<{\cal A}\right>^{3/4}$.

In both limits ${h-h_c\over\Delta'{}^{1/2}}\to 0$ and
${h-h_c\over\Delta'{}^{1/2}}\to\infty$, one recovers the same
asymptotics as the loop function found in \KoS; however, the full
loop function is different, suggesting that for loops
of the same size as a typical loop on the surface, the
physics of these two models differs qualitatively.

\subsec{$\alpha=\beta$.}
This is the line on which the critical point lies; on this
line it is possible to reexpress the function $D(h)$ using
only standard functions ($\log$, square root $\ldots$). However,
the parameters of the problem still depend on the constants
$k$, $K'$, $E'$ which are all functions of the elliptic
nome $q$. Let us see how this works.

By putting $\alpha=\beta$ we obtain from \condone\ that $x_0=K/2$ and
from \condtwo\ that
\eqn\BCA{
\eqalign{
h_1&={k' \over 4\a (1+k')}\cr
h_2&={k' \over 4 \a (1+k')}\cr
h_3&={1 \over 4 \a (1+k')^2}\cr
}}
%
%
If we employ the functional relations for $\sn$ and $\cn$, together
with the formulas
$\sn(K/2)={1\over \sqrt{k'+1}}$ etc, we find:
\eqn\DENH{
\rho(h)=1+ {2 \over i \pi }  \log {-i \sqrt{(h-h_1)(h_2-h)}+ 
\sqrt{(h_2-h_1)(h_3-h)} \over i \sqrt{(h-h_1)(h_3-h)}+ \sqrt{(h_3-h_1)(h_2-h)}  }
}
The constants $h_1$, $h_2$, $h_3$ are given by \BCA, where $\alpha$ is:
\eqn\STREQb{
\alpha={E'-(1-k')K'\over 2\pi(1+k')}
}

\appendix{B}{Small $q$ expansions.}
We rewrite the condition \condone\ as
$$
{\pi\over 2K} x_0= {\pi\over 4}+s\log q^{-1} 
$$
where it is reminded that
$$s\equiv{\log(\beta/\alpha)\over 4\pi}$$
In these variables, the defining equation \condthree\ for $\alpha$
becomes:
\eqn\COMEQ{
2\pi^2\alpha= KE'{1 \over 2s+{\theta_1'\over\theta_1}({\pi x_0\over 2K})}-
KK'{2s+{\theta_4'\over\theta_4}({\pi x_0\over 2K})\over 
\left[2s+{\theta_1'\over\theta_1}({\pi x_0\over 2K})\right]^2 \sn^2x_0} 
}
We then use the $q$ expansions:
\eqn\ELEX{\eqalign{
K &= \pi/2 (1+4q+4q^2 +O(q^3))                               \cr
K' &= {\log q^{-1} \over 2} (1+4q+4q^2 +O(q^3))               \cr
E' &= 1+ 4q \log q^{-1}-4q -8q^2 \log q^{-1}+12q^2 + O(q^3\log q^{-1})   \cr
k &= 4 q^{1/2} (1-4q+14q^2 +O(q^3))                             \cr
k' &=  1 -8q +32q^2 +O(q^3)                                     \cr
}}
and the double $q$, $s$ expansions:
\eqn\VAREX{\eqalign{
1/\sn^2(x_0) &=2(1-2s\log q^{-1}-4q+4s^2 \log^2 q^{-1} +16qs\log q^{-1}+16q^2) +\cdots \cr
{\theta_1'\over\theta_1}\left({\pi x_0\over 2K}\right) &= 1-2s \log q^{-1}+2s^2 \log^2q^{-1}+4q^2 +\cdots \cr
{\theta_4'\over\theta_4}\left({\pi x_0\over 2K}\right) & = 4q +\cdots \cr
{\theta_3'\over\theta_3}\left({\pi x_0\over 2K}\right) & = -4q +\cdots \cr
}}
where $s$ is assumed to be -- at most --
of order $q$: this turns out to be a correct assumption, since one has
$|s|\lesssim 2q$ (the limiting values define
the critical line). In \VAREX\ the error is of order $q^3 \log^3 q^{-1}$.

The deviation of $\alpha$ from criticality is then given by:
\eqn\EQSTapp{
\Delta=2s+8q^2\log q^{-1} +2s^2 \log^2 q^{-1}+4q^2 -4s^2+O(q^3 \log^3 q^{-1})
}
The analogous expression for $\Delta'=1-4\pi\beta$ is:
\eqn\EQSTappb{
\Delta'=2(1-2\pi)s+8q^2\log q^{-1} +2s^2 \log^2 q^{-1}+4q^2 -4s^2(1-2\pi+2\pi^2)+O(q^3 \log^3 q^{-1})
}
On the critical line, $s\sim \pm 2q$ and therefore
its slope is $\Delta/\Delta'\sim 1/(1-2\pi)$.

In the same way, one can expand the correlation function
\ABAB; the result is
\eqn\ABABapp{
\eqalign{
\atr{(AB)^2}=&{2\pi\over 3}
\Big(
-3+2\pi+2s(-3+9\pi-4\pi^2)-6s^2 \log^2 q^{-1}
+24\pi s^2\log q^{-1}\cr
&-24 q^2 \log q^{-1}
+4\pi s^2(3-12\pi+4\pi^2)+12 q^2(-1+4\pi)\Big)
+O(q^3 \log^3 q^{-1})\cr
}
}
which is of the form \ABABc.

\listrefs
\bye